# Unambiguous tracking of protein phosphorylation by fast high-resolution FOSY NMR

Dmitry M. Lesovoy,‡ Panagiota S. Georgoulia,‡ Tammo Diercks, Irena Matečko-Burmann, Björn M. Burmann, Eduard V. Bocharov, Wolfgang Bermel, and Vladislav Y. Orekhov*.

**ABSTRACT:** Phosphorylation is a prototypical example of post-translational modifications (PTMs) that dynamically modulate protein function, where dysregulation is often implicated in disease. NMR provides information on the exact location and time course of PTMs with atomic resolution and under nearly physiological conditions, including inside living cells, but requires unambiguous prior assignment of affected NMR signals to individual atoms. Yet, existing methods for this task base on a global, hence, costly and tedious NMR signal assignment that may often fail, especially for large intrinsically disordered proteins (IDPs). Here we introduce a sensitive and robust method to rapidly obtain only the relevant local NMR signal assignment, based on a suite of FOcused SpectroscopY (FOSY) experiments that employ the long overlooked concept of selective polarisation transfer (SPT). We then demonstrate the efficiency of FOSY in identifying two phosphorylation sites of proline-dependent glycogen synthase kinase 3 beta (GSK3β) in human Tau40, an IDP of 441 residues. Besides confirming the known target residue Ser[404], the unprecedented spectral dispersion in FOSY disclosed for the first time that GSK3β can also phosphorylate Ser[409] without priming by other protein kinases. The new approach will benefit NMR studies of other PTMs and protein hotspots in general, including sites involved in molecular interactions and conformational changes.

Post-translational modifications (PTMs) constitute an additional level of complexity in modulating protein function, where phosphorylation is the best studied signalling switch, being abundant and essential in the regulation of intrinsically disordered proteins (IDPs).[1-3] Phosphorylation is conventionally detected and quantified by mass spectrometry, then validated by mutagenesis or antibody binding assays. This procedure is expensive, time-consuming, and often fraught with problems especially for highly charged phosphopeptides, repetitive sequences, and proximal modification sites, all of which are hallmarks of IDPs.[4-5] To address these difficulties, time-resolved heteronuclear NMR methods have been developed that also allow to derive the reaction kinetics[6-10], but require prior NMR signal assignment for the entire protein. The latter is obtained from a suite of multi-dimensional experiments[11-16] with often limited sensitivity, and after several days of measurement and sophisticated spectra analysis to resolve assignment ambiguities that may still prove inextricable in highly crowded spectral regions.[17] Thus, a fast, simple, and robust NMR approach for tracking PTMs like phosphorylation would be of great benefit for elucidating their role in signalling proteins with atomic resolution.[18-20] To address this need, we here introduce FOcused SpectroscopY (FOSY) for local (instead of global) *de novo* NMR signal assignment at structural hotspots, such as PTM sites, based on a minimal set of frequency-selective experiments that focus on their sequential vicinity and combine the ultra-high signal dispersion of up to six- or seven-dimensional (6D, 7D) spectra with the sensitivity, speed, and simplicity of 2D spectra.

Central to our method of focussing onto the few residues affected by PTMs and solving the spectral dispersion problem in only two dimensions is the use of frequency **S**elective **Po**larisation **T**ransfer (SPT) to single out one coupled nuclear spin system (i.e. a residue) at a time, and with an efficiency and versatility higher than achievable by traditional broadband experiments.[21-22] While optimising sensitivity, multiple selection of known frequencies along a chosen spin system minimises spectral complexity and makes their lengthy sampling in further indirect dimensions redundant. SPT is a known technique with several clear advantages, but its use has so far been limited to isolated two-spin systems[23-25] like $^1$H-$^{15}$N amide groups, allowing a reduction by just two spectral dimensions. Thus, a 6D could be cut down to a 4D experiment that would still take impractically long to measure for each selected residue at a time. Here we introduce frequency **S**elective and **S**pin-**S**tate **S**elective **P**olarisation **T**ransfer (S$^4$PT) as a generalized SPT approach for coupled multi-spin systems, as in isotopically labelled amino acids, which allows to eliminate several pertaining spectral dimensions as well as detrimental evolution of competing passive spin couplings. The novel 2D FOSY spectra (Figure 1) yield a signal dispersion as in a 6D HNCOCANH[11] and 7D HNCOCACBNH, yet with far superior sensitivity and ease of analysis.

As a showcase application for the proposed FOSY approach to monitor PTMs in IDPs, we identify phosphorylated residues *in vitro* by proline-dependent glycogen synthase kinase 3 beta (GSK3β)[26] in the 441-residue long human hTau40 protein, which comprises 80 serines and threonines as potential phosphorylation sites. Abnormal hTau40 hyper-phosphorylation is directly linked to dysregulation and, possibly, aggregation that characterises neurodegenerative tauopathies such as Alzheimer's disease.[27-31] As a first step of our strategy (outlined in Figure 2a), spectral changes from GSK3β-mediated phosphorylation of hTau40 in the most sensitive and best dispersed 3D HNCO spectrum revealed several shifted or newly appearing signals for p-hTau40 (peaks a–g in Figure S4). Due to the abundance of proline and glycine residues in IDPs, we assembled a list of hTau40 sequence stretches conforming with the general motif (P/G)-$X^n$-p(S/T)-$X^n$-(P/G) where X ≠ Pro, Gly. The list can be further refined based on reported phosphorylation sites or known kinase consensus sequence motifs (Table S1). This preparatory step concludes with the acquisition of two complementary proline selective 2D experiments[32] to identify the residues following (PX-) or preceding (-XP) prolines (Figure S5).

To start the FOSY assignment process, we focus on signal 0 (Figure 2b) that appears in the 3D HNCO spectrum after hTau40 phosphorylation (peak 'a' in Figure S4) and likely corresponds to a phosphorylated serine (pS) or threonine (pT). The proline selective spectra show that this new signal derives from a pS or pT preceding a proline (i.e., a p(S/T)P motif, Table S1). A pair of 2D FOSY hnco(CA)NH and hncoCA(N)H experiments (Figure 1) is then recorded for signal 0 using its associated exact $^1H^{N,0}$, $^{15}N^{H,0}$, $^{13}C^{O,-1}$ frequencies from the 3D HNCO spectrum. This yields the $^{15}N^{H,-1}$, $^1H^{N,-1}$, and $^{13}C^{A,-1}$ frequencies of the preceding residue X and connects signal 0 with signal -1 (Figure 2b) to compose a -Xp(S/T)P motif. We continue the FOSY walk to signal -2 using the $^{15}N^{H,-1}$, $^1H^{N,-1}$, and $^{13}C^{O,-2}$ frequencies for signal -1 (the latter again derived from the 3D HNCO). This iteration is repeated until reaching a proline or glycine signal, the latter identified by the negative intensity (from constant-time evolution) and characteristic chemical shift of its $^{13}C^A$ signal. For starting signal 0 we, thus, identify a GXXp(S/T)P motif that matches with either GYSS$^{199}$P or GDTS$^{404}$P stretch in the hTau40 sequence.

To resolve this assignment ambiguity, or in case of several peaks observed in the 2D hnco(CA)NH and hncoCA(N)H spectra due to an overlap of the initially selected frequencies (e.g., the FOSY walk from signal -2 connects with both signals -3 and -3* in Figure 2b), we furthermore record a 2D FOSY hncocacbNH experiment (Figure 1) with five fixed frequencies ($^1H^{N,0}$, $^{15}N^{H,0}$, $^{13}C^{O,-1}$, $^{13}C^{A,-1}$, $^{13}C^{B,-1}$) to determine the preceding residue type. The unknown $^{13}C^{B,-1}$ frequency to be tested is chosen from published residue-type-specific chemical shifts for random coil.[33-34] By producing a signal only if the correct $^{13}C^{B,-1}$ frequency is used for selective decoupling[35-36], the 2D FOSY hncocacbNH spectrum for signal 0 reveals that the associated p(S/T) residue is preceded by a threonine. Thus, GDTS$^{404}$P is the correct assignment and signal 0 corresponds to pS$^{404}$.

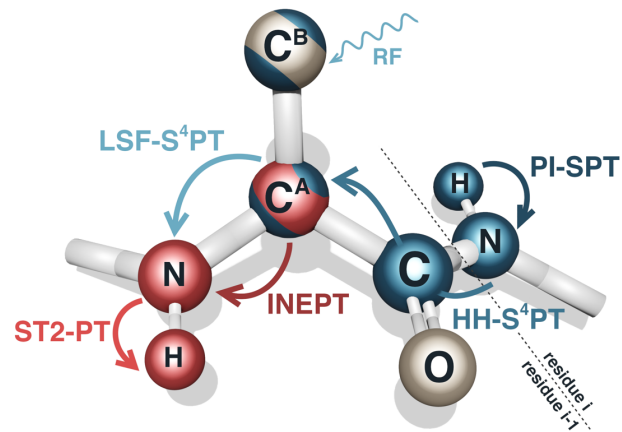

**Figure 1.** Polarisation transfer pathway in the 2D FOSY hnco(CA)NH, hncoCA(N)H, and hncocacbNH experiments (see Supplementary Figure S1). Colour code for polarisation transfer steps shown by arrows: blue – frequency selective (PI-SPT, HH-S$^4$PT, LSF-S$^4$PT), red - broadband (INEPT, ST2-PT[32, 37]). Colour code for atoms according to their frequency probing: blue – known (e.g., from a 3D HNCO) and selected by SPT; red – unknown and evolved in a spectral dimension; striped red/blue – optionally selected or evolved; grey - ignored. Transfer starts with amide proton $^1H^i$ magnetization of residue $i$ and ends with detection on $^1H^{i-1}$ of the preceding residue $i-1$. As key element of FOSY experiments, we introduce frequency selective and spin-state selective polarization transfer (S$^4$PT) steps (detailed in the Supplementary Information) that can be specifically optimized for local spin system properties like relaxation[21], chemical[22] or conformational[23-24] exchange, scalar coupling network, and chemical shift pattern. The experiments start with selective polarisation transfer by population inversion (PI-SPT) of the TROSY[37] component of $^1H^i$ magnetization to create $2H_z^i N_z^i$ anti-phase polarisation with a maximal efficiency surpassing all methods for broadband polarisation transfer.[21] A subsequent HH-S$^4$PT step implements separate selective heteronuclear Hartmann-Hahn transfer for both $^{15}N$ TROSY and anti-TROSY coherences without their mixing, as required by the TROSY principle[37], to achieve relaxation optimised fast and direct $2H_z^i N_z^i \rightarrow 2C_z^{A,i-1} C_z^{O,i-1}$ conversion. A final LSF-S$^4$PT (instead of broadband INEPT) step, used only in the FOSY hncocacbNH experiment, employs longitudinal single field polarization transfer[38] for direct $2C_z^{A,i-1} C_z^{O,i-1} \rightarrow 2C_x^{A,i-1} N_z^{i-1}$ conversion and concomitant selective $^{13}C^{B,i-1}$ decoupling to probe for its amino acid type specific frequency. All FOSY experiments ensure maximal preservation of both water and aliphatic proton polarization to enable fast selective polarization recovery[11] for the amide protons.

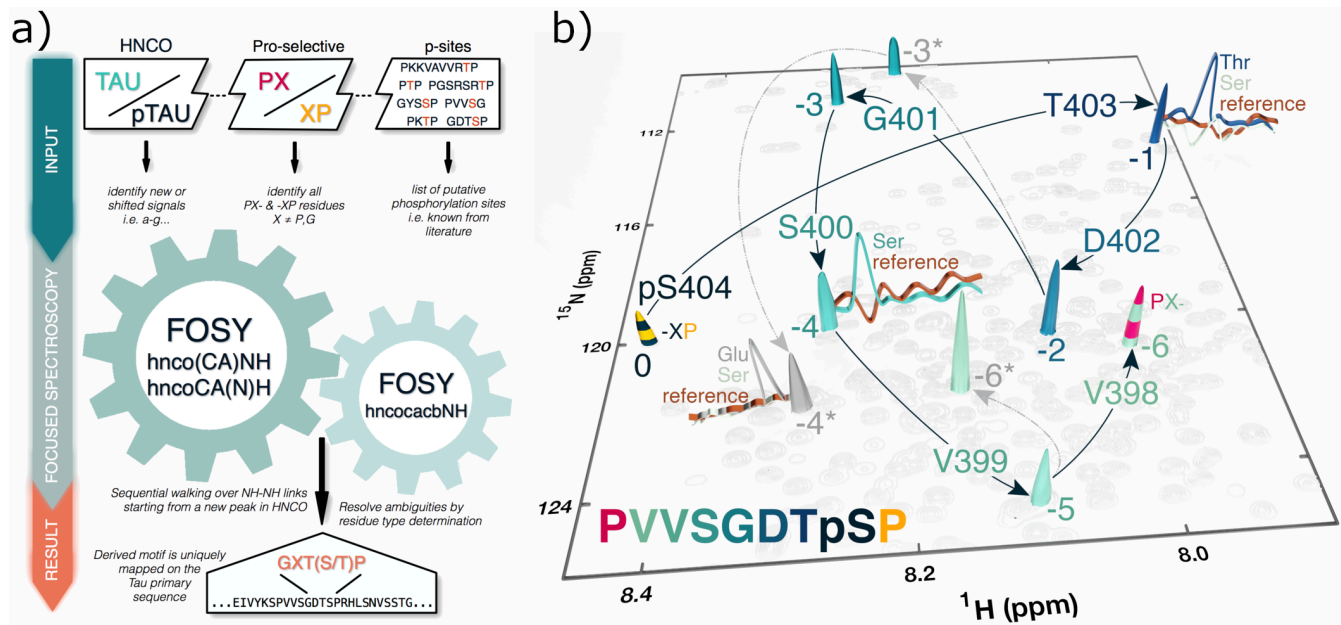

**Figure 2.** FOSY NMR assignment strategy **(a)** and application **(b)** to assign a phosphorylation site in hTau40. Signal 0 newly appears after hTau40 phosphorylation by GSK3β and corresponds to a residue preceding a proline, as revealed by a proline-selective experiment. The sequential walk (black arrows) along NH-NH correlations is traced out by successive iterations of 2D FOSY-hnco(CA)NH experiments. Thus connected signals are numbered by the pertaining FOSY step number, and are gradually coloured as in the associated peptide sequence (below, left). For signal -2, the 2D FOSY-hnco(CA)NH spectrum also opens an alternative branch of signals indicated by asterisks and connected by dashed grey arrows. To resolve such ambiguities, preceding residue types were tested using the 2D FOSY-hncocacbNH experiment, which only produces a signal (inserted 1D $^1$H projections) if the correct residue specific $^{13}C^{\beta,-1}$ frequency is preset. The derived PXXSGXTp(S/T)P motif unambiguously maps to PVVSGDTS$^{404}$P and, thus, assigns the phosphorylation site signal 0 to pS$^{404}$.

Similar 2D FOSY hncocacbNH experiments for the alternative Gly signals -3 and -3* show that only the former is preceded by a Ser (signal -4), as in the identified GDTS$^{404}$P stretch, while signal -3* is preceded by a Glu (signal -4*) and, therefore, starts a false branch. Further FOSY walking leads to signals -5 and -6 that the proline-selective spectrum shows to succeed a proline (in contrast to signal -6* on another false branch). This results in a PXXSGXTp(S/T)P motif that uniquely maps onto the hTau40 PVVSGDTS$^{404}$P stretch and, thus, corroborates signal 0 assignment to pS$^{404}$. Overall, the minimal set of 2D FOSY spectra (Figure S5) for unambiguous identification of the pS$^{404}$ site comprises three pairs of complementary hnco(CA)NH and hncoCA(N)H for the sequential walk plus three hncocacbNH to identify the preceding residue type, altogether recorded within less than two hours. As these highly selective 2D spectra contain only a single or very few peaks, their real-time analysis is most straightforward. Analogous identification of the second phosphorylation site pS$^{409}$ by a similar minimal set of 2D FOSY spectra to compile the unique PRHLpS$^{409}$ motif is illustrated in the Supporting Information (Figure S6).

Together, pS$^{404}$ and pS$^{409}$ account for all newly appearing and shifted signals (of nearby residues) in the HNCO spectrum of p-hTau40, proving them to be the only two phosphorylation sites for GSK3β under our experimental conditions. Of note, while proline dependent S$^{404}$ phosphorylation was known from prior studies,[26, 39] only the extreme spectral simplification and dispersion afforded by our new FOSY approach allowed to also confirm S$^{409}$ phosphorylation by GSK3β. The latter was previously suggested to occur in the pre-neurofibrillar tangle state of hTau40,[30] but so far could not be unambiguously shown to occur without priming by other protein kinases.[40]

In summary, we have demonstrated the *de novo* identification of phosphorylation sites in IDPs by the fast and robust new FOSY NMR approach. The proposed strategy requires no lengthy prior signal assignment nor knowledge of the target sites for PTM. Key to the new approach is its local focus on the relevant modification sites to identify short motifs for unambiguous sequence mapping, which is much faster and broader applicable than the conventional approach of global NMR signal assignment limited by the size and spectral complexity of the protein. Analysis is fast and most straightforward due to the extreme simplicity of 2D FOSY spectra that still reflect the enormous signal dispersion of their conventional complex 6D and 7D counterparts. The great benefits and versatility of such frequency selective NMR approaches, here demonstrated by a proof-of-concept study on IDP phosphorylation, are generalizable and should enable broad new studies on a plethora of biomolecular hotspots hitherto inaccessible by NMR.

## AUTHOR INFORMATION

Corresponding Author

**Vladislav Y. Orekhov** - Department of Chemistry and Molecular Biology, University of Gothenburg, Box 465, 405 30, Gothenburg, Sweden; Swedish NMR Centre, University of

Gothenburg, Box 465, 405 30, Gothenburg, Sweden. ORCID: 0000-0002-7892-6896

Email: vladislav.orekhov@nmr.gu.se

Author Contributions

‡These authors contributed equally.

Authors

**Dmitry M. Lesovoy** - Department of Structural Biology, Shemyakin-Ovchinnikov Institute of Bioorganic Chemistry RAS, 117997, Moscow, Russia; Research Center for Molecular Mechanisms of Aging and Age-related Diseases, Moscow Institute of Physics and Technology (State University), 141700 Dolgoprudny, Russia. ORCID: 0000-0002-9130-715X

**Panagiota S. Georgoulia** - Department of Chemistry and Molecular Biology, University of Gothenburg, Box 465, 405 30, Gothenburg, Sweden ORCID: 0000-0003-4573-8052

**Tammo Diercks** – NMR Platform, CiC bioGUNE, Bld. 800, Parque Tecnológico de Bizkaia, 48160 Derio, Spain. ORCID: 0000-0002-5200-0905

**Irena Matečko-Burmann** - Department of Psychiatry and Neurochemistry, University of Gothenburg, 405 30 Gothenburg, Sweden; Wallenberg Centre for Molecular and Translational Medicine, University of Gothenburg, 405 30 Gothenburg, Sweden. ORCID: 0000-0001-8873-8381

**Björn M. Burmann** - Department of Chemistry and Molecular Biology, University of Gothenburg, Box 465, 405 30, Gothenburg, Sweden; Wallenberg Centre for Molecular and Translational Medicine, University of Gothenburg, 405 30 Gothenburg, Sweden. ORCID: 0000-0002-3135-7964

**Eduard V. Bocharov** - Department of Structural Biology, Shemyakin-Ovchinnikov Institute of Bioorganic Chemistry RAS, 117997, Moscow, Russia; Research Center for Molecular Mechanisms of Aging and Age-related Diseases, Moscow Institute of Physics and Technology (State University), 141700 Dolgoprudny, Russia. ORCID: 0000-0002-3635-1609

**Wolfgang Bermel** - Bruker BioSpin GmbH, Silberstreife, 76287 Rheinstetten, Germany.

**Funding Sources**

The work was sponsored by Swedish Research Council to V.Y.O. (grant 2019-03561) and Russian Science Foundation to D.M.L. and E.V.B. (grant 19-74-30014, in part of NMR pulse sequence design). B.M.B. gratefully acknowledges funding from the Knut och Alice Wallenberg Foundation through the Wallenberg Centre for Molecular and Translational Medicine, University of Gothenburg, Sweden.

**Notes**

The presented pulse sequences for Bruker Avance spectrometers are available at https://github.com/lesovoydm/FOSY-NMR.

## ACKNOWLEDGMENT

The authors thank Dr. Ilya S. Kuprov for helpful discussions.

# Supporting Information

# Unambiguous tracking of protein phosphorylation by fast, high-resolution FOSY NMR.

Dmitry M. Lesovoy,[‡] Panagiota S. Georgoulia,[‡] Tammo Diercks, Irena Matečko-Burmann, Björn M. Burmann, Eduard V. Bocharov, Wolfgang Bermel, and Vladislav Y. Orekhov*

[‡] *These authors contributed equally.*

* Corresponding author. e-mail: vladislav.orekhov@nmr.gu.se

# Selective Polarisation Transfer schemes

The FOSY experiments use a concatenation of highly selective polarisation transfer (SPT) steps to single out a specific coupled spin system, i.e. residue, by selecting several of its known nuclear frequencies. Various schemes for SPT have been reported that differ, e.g. in transfer mechanism, transfer time, achievable selectivity, susceptibility to relaxation and competing passive couplings (given by the local spin topology). Notably the latter demands vary significantly along the desired polarisation transfer pathway through the peptide spin system. To maximise efficiency and robustness, FOSY therefore implements and adjusts the respectively best suited SPT scheme for each required transfer step. The key for concatenation of the SPTs into the experiment polarisation flow is the newly introduced frequency selective and spin state selective polarisation transfer S$^4$PT, as described below.

*PI-SPT*

For a heteronuclear IS two spin-½ system with mutual scalar $J_{IS}$ coupling, I spin coherence produces an in-phase signal doublet with frequencies $v^I + J_{IS}/2$ and $v^I - J_{IS}/2$ as the coupled S spin may be aligned either parallel ($S_\alpha$) or antiparallel ($S_\beta$) with the external magnetic field, respectively. An inversion pulse applied with high selectivity at the frequency $v^I - J_{IS}/2$ then creates $\gamma_S$ enhanced anti-phase $2I_zS_z = I_zS_\alpha - I_zS_\beta$ magnetization. This scheme for magnetization transfer,[1-2] which we abbreviate PI-SPT (Population Inversion for Selective Polarisation Transfer), is frequency selective for spin I and notably immune to any further (other than $J_{IS}$) passive scalar couplings of spin S. Furthermore, in proteins, the amide HN group forms a particular IS system with significant cross-correlated relaxation giving rise to a strong TROSY effect.[3] For such systems it was shown that PI-SPT reaches the physical limit of efficiency for magnetization transfer, surpassing all broadband PT schemes, if the $H_zN_\beta$ TROSY polarisation component is selectively inverted by a CROP-shaped pulse[4]. In FOSY experiments, we employ such optimal PI-SPT for the initial $H_z \rightarrow 2N_zH_z$ transfer step.

For the following section, we note that term $2I_zS_z$ can be also presented as $2I_zS_z = I_\alpha S_z - I_\beta S_z$, where $I_{\alpha|\beta}$ present single spin states of spin I. Both $I_\alpha S_z$ and $I_\beta S_z$ terms are directly used by the subsequent S$^4$PT block without need for intermittent complete magnetization conversion of $2I_zS_z$ to spin S.

*HH-S$^4$PT - Heteronuclear Hartmann-Hahn (HH) frequency Selective and Spin-State Selective Polarisation Transfer (S$^4$PT)*

In an isolated heteronuclear IS two-spin system with mutual $J_{IS}$ coupling, selective I→S Hartmann-Hahn polarization transfer[5] (HH-SPT) can be achieved by simultaneous weak continuous wave (CW) irradiation at the exact $v_I$ and $v_S$ frequencies for a duration $\tau_{CW} = 1/J_{IS}$ and with identical field strengths

$$B_{1,I} = B_{1,S} = \frac{J_{IS}\sqrt{4n^2-1}}{4} \quad [S1]$$

where $n$ = 1, 2, etc. and $n$ = 1 provides the weakest radiofrequency field $B_1 = J_{IS} \cdot \sqrt{3}/4$, affording highest $v_I$ and $v_S$ frequency selectivity as well as maximal tolerance to relative $B_1$ miscalibration and mismatch. These latter benefits of frequency selective heteronuclear Hartmann-Hahn transfer are crucially important and in stark contrast to the broadband implementation of Hartmann-Hahn transfer, where the losses from $B_1$ mismatch (that scale with $B_1$ and are inevitable for separate probe coils for I and S) are exacerbated and effectively prevent a wider use in solution state NMR.

Further passive couplings $J_{SM}$ of the receiving spin S, however, may strongly compromise the efficiency of HH-SPT. If $J_{SM} \approx J_{IS}$, the detrimental effects can be efficiently suppressed by simply using a slightly stronger $B_1$ field, given by Eq. S1, for $n$ = 2 or 3, in exchange for a small loss of frequency selectivity in the Hartmann-Hahn polarisation transfer. If $J_{SM} > J_{IS}$, however, the signal of spin S gets "broadened" beyond the extremely narrow HH-SPT bandwidth of approximately $B_1 \approx J_{IS}$ (see above) due to splitting by $J_{SM}$. Efficient HH-SPT then requires to separately irradiate both lines of the S signal doublet[1-2, 6] with the same weak $B_1$ field strength given by Eq. S1, i.e. CW irradiation at the three frequencies $v_I$, $v_S - J_{SM}/2$, and $v_S + J_{SM}/2$. Under these conditions, HH-SPT passes via separate parallel $I_x \rightarrow S_xM_\alpha$ and $I_x \rightarrow S_xM_\beta$ pathways that are immune to $J_{SM}$ coupling evolution and can each be described by the HH-SPT formalism for an isolated two spin system.[5] As the parallel heteronuclear Hartmann-Hahn polarisation transfer pathways are both ($v_I$, $v_S$) frequency selective and ($M_{\alpha|\beta}$) spin state selective, we propose the acronym HH-S$^4$PT. Importantly, the sign for $I_x \rightarrow S_xM_\alpha$ and $I_x \rightarrow S_xM_\beta$ pathways in HH-S$^4$PT can be controlled via the phase of the pertaining CW irradiation at $v_S - J_{SM}/2$ vs. $v_S + J^{SM}/2$. Thus, identical CW phases produce $S_xM_\alpha + S_xM_\beta = S_x$ inphase transfer while opposite phases $S_xM_\alpha - S_xM_\beta = 2S_xM_z$ achieve antiphase transfer. Furthermore, HH-S$^4$PT can be generalized to larger than three-spin systems, where passive couplings "broaden" both spin I and S resonances, by adjusting the number of CW irradiation frequencies.

In FOSY experiments, we employ HH-S$^4$PT for direct $2N_x^i H_z^i \rightarrow 2CO_x^{i-1} CA_z^{i-1}$ antiphase-to-antiphase polarisation transfer. Both the starting N and receiving CO spins show passive couplings ($^1J_{N,H} \approx 90$ Hz and $^1J_{CO,CA} \approx 55$ Hz) much larger than the active $^1J_{N,CO} \approx 15$ Hz coupling. Consequently, HH-S$^4$PT must enable the following four transfer pathways:

$$N_x^i H_\alpha^i \rightarrow CO_x^{i-1} CA_\alpha^{i-1}$$
$$N_x^i H_\alpha^i \rightarrow -CO_x^{i-1} CA_\beta^{i-1}$$
$$-N_x^i H_\beta^i \rightarrow CO_x^{i-1} CA_\alpha^{i-1}$$
$$-N_x^i H_\beta^i \rightarrow -CO_x^{i-1} CA_\beta^{i-1}$$

to sum up to the desired overall transfer:

$$2N_x^i H_z^i = N_x^i(H_\alpha^i - H_\beta^i) \xrightarrow{HH-S4PT} CO_x^{i-1}(CA_\alpha^{i-1} - CA_\beta^{i-1}) = 2CO_x^{i-1} CA_z^{i-1}$$

This is achieved by CW irradiation with identical $B_1$ (given by Eq. S1) at the four frequencies $v_N^i + \frac{J_{HN}}{2}$, $v_N^i - \frac{J_{HN}}{2}$, $v_{CO}^{i-1} + \frac{J_{COCA}}{2}$, $v_{CO}^{i-1} - \frac{J_{COCA}}{2}$ and with inverted irradiation phases at $v_N^i - \frac{J_{HN}}{2}$ and $v_{CO}^{i-1} - \frac{J_{COCA}}{2}$ (see Figures S2b and c). Since the nitrogen CWs are applied at two different frequencies, it is important to align their phases at the beginning. Similarly, the two carbon CWs are phase-aligned at the end.

*LSF-S$^4$PT*

Similar to the PI-SPT described above, for an IS spin system, the Longitudinal Single Field Selective Polarization Transfer module[5, 7] (LSF-SPT) uses a single radiofrequency and thus is frequency selective only for spin I. The LSF-SPT does not depend on frequency and passive couplings of spin S and it performs efficient transfer of spin I magnetization to antiphase coherence of spin I in respect to spin S ($I_z \rightarrow 2I_xS_z$). The LSF-SPT uses a continuous-wave irradiation of spin I at frequency $v_I$ with optimal duration $\tau_{CW} = \frac{\sqrt{2}}{2J_{IS}}$ and strengths $B_1 = \frac{J_{IS}}{2}$. As for HH-SPT, the use of CW irradiation in LSF-SPT allows S$^4$PT implementation to remove detrimental effects of passive scalar couplings of the starting I spin. Thus, for a three spin ISM system with passive couplings $J_{MS}=0$ and $J_{MI} > J_{IS}$, the LSF-S$^4$PT with the $\tau_{CW}$ and $B_1$ defined above is performed without decoupling of spin M by two CW's at frequencies $v^I - J_{IM}/2$ and $v^I + J_{IM}/2$. Transfers $I_z \rightarrow 2I_xS_z$ and $2M_zI_z \rightarrow 2I_xS_z$ are achieved using the CW's aligned at the start with the same and opposite phases, respectively. The LSF-S$^4$PT approach can be extended to more passive scalar couplings of spin I by combining the corresponding synphase and antiphase CW's.

In contrast, small passive scalar couplings $J_{MI} < J_{IS}$ of spin I are handled by adjusting both $\tau_{CW}$ and $B_1$. For the desired $2C_z^{A,i-1} C_z^{O,i-1} \rightarrow 2C_x^{A,i-1} N_z^{i-1}$ transfer by active $^1J_{CA,i-1N,i-1} \approx 12$ Hz in the presence of passive $^2J_{CA,i-1N,i} \approx 7$ Hz coupling, setting $\tau_{CW} = 43$ ms and $B_1 = 8$ Hz yields 72% transfer efficiency (disregarding relaxation) similar to broadband INEPT (compare with round parentheses in Fig. S1a). The competing smaller $^2J_{CA,i-1N,i}$ coupling results in some back-transfer to $2C_x^{A,i-1} N_z^i$ about three times less efficient than the desired transfer via $^1J_{CA,i-1N,i-1}$. Unlike broadband INEPT, LSF-SPT would enable even 100% transfer when selectively decoupling the $^{15}N^i$ frequency. Yet, this may be problematic if the $^{15}N^{i-1}$ and $^{15}N^i$ frequencies are close and we, therefore, have not implemented $^{15}N^i$ decoupling in our experiments.

# FOSY experiments

The 2D FOSY $^1$H-$^{15}$N hnco(CA)NH and $^1$H-$^{13}$C$^A$ hncoCA(N)H experiments for deuterated proteins (Figure S1a) start with highly selective population inversion (PI-SPT, Figure S2a) to create $v^H$-frequency selected and $\gamma_H$ enhanced $2H_z^i N_z^i$ two-spin magnetization. Then, a hard 90° pulse on $^{15}$N produces anti-phase coherence $2H_z^i N_x^i$ for the subsequent selective Hartmann-Hahn polarisation transfer (HH-S$^4$PT) step $2H_z^i N_x^i \rightarrow 2C_x^{O,i-1} C_z^{A,i-1}$. Then, chemical shifts of $^{13}$C$^{A,i-1}$ or $^{15}$N$^{i-1}$ are traditionally sampled during corresponding constant time periods followed by TROSY $^1$H$^{i-1}$ detection. With two unknown frequencies (H$^{i-1}$ and either N$^{i-1}$ or CA$^{i-1}$) sampled and three known frequencies (H$^i$, N$^i$, CO$^{i-1}$) preselected, a pair of the 2D FOSY hnco(CA)NH and hncoCA(N)H spectra corresponds to a 3D FOSY hncoCANH (with both CA$^{i-1}$ and N$^{i-1}$ frequency sampling) that has the frequency dispersion of a pertaining 6D experiment sampling all six nuclei.

In the 2D FOSY hncocacbNH experiment (Figure S1c), INEPT transfer from C$^{A,i-1}$ to N$^{i-1}$ is replaced by the LSF-S$^4$PT module *qcw6* (Figure S2d) for direct $2C_z^{O,i-1} C_z^{A,i-1} \rightarrow 2C_x^{A,i-1} N_z^{i-1}$ transfer with concomitant residue type selective C$^{B,i-1}$ decoupling[15], as explained above. The additional frequency selection of C$^{A,i-1}$ and C$^{B,i-1}$ here produces an effective frequency dispersion equivalent to a 7D experiment. The employed field strength of the C$^{B,i-1}$ decoupling of several hundred Hertz provides only relatively low resolution in the C$^{B,i-1}$ dimension, but it is sufficient in most cases to identify the preceding residue type. To cancel out any perturbation of C$^{A,i-1}$ coherence, we compensate the effects of C$^{B,i-1}$ decoupling at the opposite side of the $^{13}$C$^{A,i-1}$ resonance (Figure S2d). If the C$^{B,i-1}$ frequency is unknown, the corresponding $v_{CA}^{i-1}$ frequency "splitting" by $^1J_{CA,CB}$ coupling is handled in the same way as the other passive $^1J_{CA,CO}$ coupling, i.e. by implementing the S$^4$PT module *qcw6* with quadruple frequency selective CW irradiation (Figure S2e), as explained above.

For non-deuterated proteins, broadband $^1$H decoupling in the hnco(CA)NH and hncoCA(N)H experiments (Figure S1a) is implemented by a pair of $^1$H inversion pulses (Figure S3a) that also ensures a return of both water and aliphatic proton magnetisations towards their thermal equilibrium. To similarly adapt the hncocacbNH experiment, we replace the module shown in Figure S1c by the block depicted in Figure S3b. The corresponding LSF-S$^4$PT sandwiches with and without $^{13}$C$^{B,i-1}$ decoupling are depicted in Figure S3c and Figure S3d, respectively. Of note, as the LSF-S$^4$PT scheme employs irradiation only on $^{13}$C, it naturally leaves all (water and protein) $^1$H magnetization unaffected.

For IDPs (like Tau) with their typically slow transverse relaxation, the $^{15}$N$^i$ frequency selectivity can be further improved as shown in Figure S1b. Here, the initial PI-SPT module is replaced by a selective 90° pulse on $^1$H$^i$ followed by HH-SPT employing simultaneously applied CW irradiation modules *scw1* (on H$^i$) and *scw2* (on N$^i$) for $H_x^i \rightarrow N_x^i$ magnetisation transfer. The directly appended HH-S$^4$PT step for $N_x^i \rightarrow 2C_x^{O,i-1} C_z^{A,i-1}$ is implemented by single $v_N^i$ frequency selective (*scw3*) CW irradiation on N$^i$ with concomitant selective H$^i$ decoupling by *scw5*. The use of *scw3* (Figure S1b) instead of double $v_N^i \pm \frac{J_{HN}}{2}$ frequency selective irradiation *dcw3* (Figure S1a, S2b) results in a twice as high $v_N^i$ frequency selectivity, while the $v_{CO}^{i-1}$ frequency selectivity of the same *dcw4* module (Figure S2c) is maintained. The block shown in Figure S1b can be used both for deuterated and non-deuterated proteins, and illustrate the versatility of the FOSY approach along with its possible fine tuning to a specific spin system. In this case, while the PI-SPT version of the experiment (Figure S1a) minimises relaxation losses and, thus, maximises sensitivity, its HH-SPT version (Figure S1b) optimises for frequency selectivity and effective spectral dispersion.

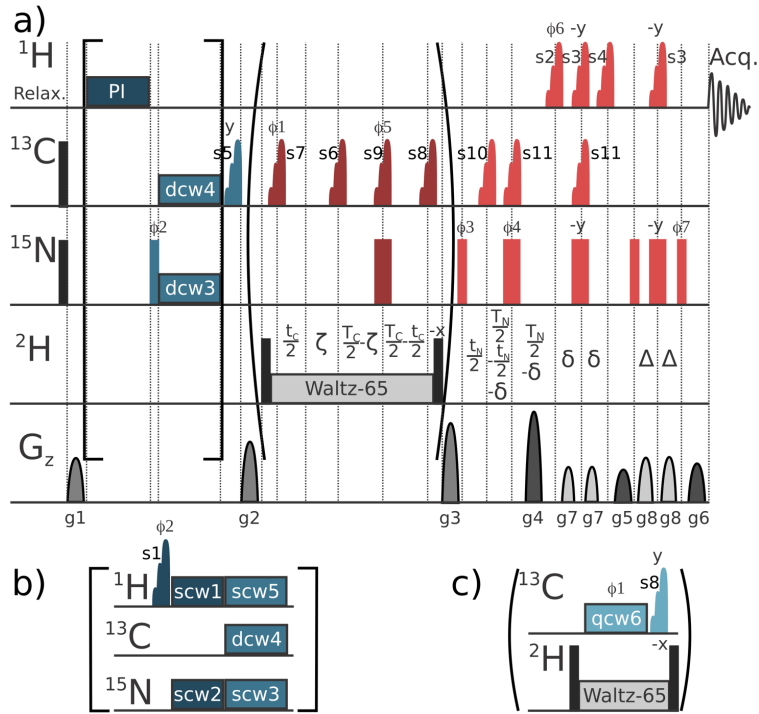

**Figure S1.** Inter-residual FOSY experiments for deuterated proteins in H$_2$O-based buffer solution. Selective (PI-SPT, HH-S$^4$PT, LSF-S$^4$PT) and broadband (INEPT, ST2-PT) polarisation transfer blocks are colored in blue and red, respectively, same as in Figure 1 of the article. a) The 2D FOSY $^1$H-$^{15}$N hnco(CA)NH and $^1$H-$^{13}$C$^A$ hncoCA(N)H experiments are acquired as the corresponding 2D planes of 3D FOSY-hncoCANH experiment depicted in the panel. Initial selective population inversion (PI) of the TROSY component by a frequency selective pulse (applied at $\nu_H^i - \frac{J_{HN}}{2}$; see Fig. S2a) produces $2H_z^i N_z^i$ antiphase magnetisation[1-2, 4] that is converted into $2H_z^i N_x^i$ coherence by a (unselective) 90° $^{15}$N pulse. The following HH-S$^4$PT block with a double continuous-wave (dcw3, see Fig. S2) simultaneously transfers both $^{15}$N TROSY and anti-TROSY components without proton decoupling, i.e. the transfer (selective for $^{15}$N$^{H,i}$ and $^{13}$C$^{O,i-1}$) $2H_z^i N_x^i \to 2C_x^{O,i-1} C_z^{A,i-1}$ is achieved by a 66.7 ms HH-S$^4$PT block including: double selective at frequencies $^{15}$N$^{H,i}$ ±46 Hz continuous waves (dcw3) with $B_{dcw3}$ = 14.5 Hz beginning with aligned phases ±x; as well as double selective at frequencies $^{13}$C$^{O,i-1}$ ±28 Hz continuous waves (dcw4) with the same $B_{dcw4}$ = 14.5 Hz ending the phases aligned ±x. Constant-time States-TPPI evolution of $^{13}$C$^{A,i-1}$ (including broadband $^2$H$^A$-decoupling) is implemented with $t_C$ delay and φ1. The subsequent INEPT step (in brackets) for simultaneous evolutions $2C_z^{O,i-1} C_y^{A,i-1} \to C_x^{A,i-1}$ (with $2\zeta = (2 \cdot ^1J_{CACO})^{-1}$ = 9.1 ms) and $C_x^{A,i-1} \to 2C_y^{A,i-1} N_z^{i-1}$ (during $T_C < (2 \cdot ^1J_{CAN})^{-1} \approx 28.6$ ms) is implemented with $^2$H$^A$ decoupling and optional States-TPPI type constant-time sampling of the $^{13}$C$^{A,i-1}$ frequency (during $t_C \le T_C$). The following INEPT step for $2C_z^{A,i-1} N_y^{i-1} \to N_x^{i-1}$ transfer (during $T_N \le (2 \cdot ^1J_{CAN})^{-1} \approx 40$ ms) is concatenated with the final ST2-PT module and implemented with optional echo/anti-echo type constant-time sampling of the N$^{i-1}$ frequency (during $t_N \le T_N - 2\delta$) with simultaneous alteration of three gradients g4*EA4, g5*EA5, g6*EA6 (EA4 = 1, 0.8750; EA5 = 0.6667, 1; EA6 = 1, 0.6595) and φ6 = y, -y, φ7 = y, -y. The constant-time delays $T^C$ and $T^N$ are 14.3 ms and 20 ms respectively; delays are $\zeta = 1/4 \, ^1J_{CACO}$ = 4.7 ms and $\delta \approx \Delta \approx 1/4 \, ^1J_{NH}$ = 2.75 ms. Pulses: Narrow and thick bars represent 90° and 180° high power pulses, respectively. Selective pulses[4] on amide $^1$H: s1(90°) = 1 ms Sinc at the selected $\nu_H^i$; s2(90°) = 1.74 ms time reversed EBurp2, s3(180°) = 1.85 ms ReBurp, and s4(90°) = 1.74 ms EBurp2 are applied at 8.5 ppm (centre of amide $^1$H). Selective pulses on $^{13}$C: s5(90°) = 600 µs Sinc and s6(180°) = 600 µs Sinc at the selected $\nu_{CO}^{i-1}$; s7(90°) = 500 µs Q5 and s8(90°) = 500 µs time reversed Q5 at 55 ppm (center of CA); s9(180°) = 200 µs Q3 at 39 ppm (centre of CA and CB); s10(180°) = 800 µs IBurp1 at 173 ppm (centre of CO); s11(180°) = 600 µs IBurp1 at 55 ppm (centre of CA). Gradients: All gradient pulses have smoothed square (SMSQ) shape and 1ms duration except for g4, g5, g6 (0.5 ms) and g7, g8 (0.3 ms). Relative gradient strengths: g1 = 57%, g2 = 47%, g3 = 37%, g4 = -80%, g5 = 30%, g6 = 30.13%, g7 = 15%, g8 = 60%, whereas 100% corresponds to 53.5 G/cm. Pulse phases (default = x): φ1=x,-x; φ2=2(-y), 2(y); φ3=4(x), 4(-x); φ4=8(x), 8(-x); φ5=16(x), 16(-x); φ6=y; φ7=y; φrec=4(x, -x, -x, x, -x, x, x, -x). b) Modification to double the $\nu_N^i$ selectivity. The indicated block replaces the pulse program segment in square brackets in panel (a). The initial PI is replaced by a $\nu_H^i$ selective s1(90°) pulse followed by a HH-SPT block ($\tau_{scw1/2}$ = 11 ms) employing simultaneous single frequency selective CW irradiation at $\nu_H^i$ (scw1) and $\nu_N^i$ (scw2) with identical strength $B_{scw1} = B_{scw2}$ = 39 Hz. The subsequent dcw3 at $\nu_N^i$ is replaced by single frequency CW (scw3) with maintained $B_{scw3}$ = 14.5 Hz and concomitant selective CW decoupling (scw5) at $\nu_H^i$ with $B_{scw5}$ = 700 Hz. ). c) 2D FOSY-hncocacbNH with fixation of $\nu_{CA}^{i-1}$ and $\nu_{CB}^{i-1}$ frequencies. The concatenated INEPT (round brackets, in panel a) is replaced by a LSF-S$^4$PT module ($\tau_{qcw6}$ = 59 ms) employing quadruple frequency selective CW (qcw6; see Fig. S2d) at $\nu_{CA}^{i-1} \pm 28 Hz$ with $B_1$ = 6 Hz and the frequency selective decoupling of $^{13}$C$^{B,i-1}$ with corresponding compensation irradiation.

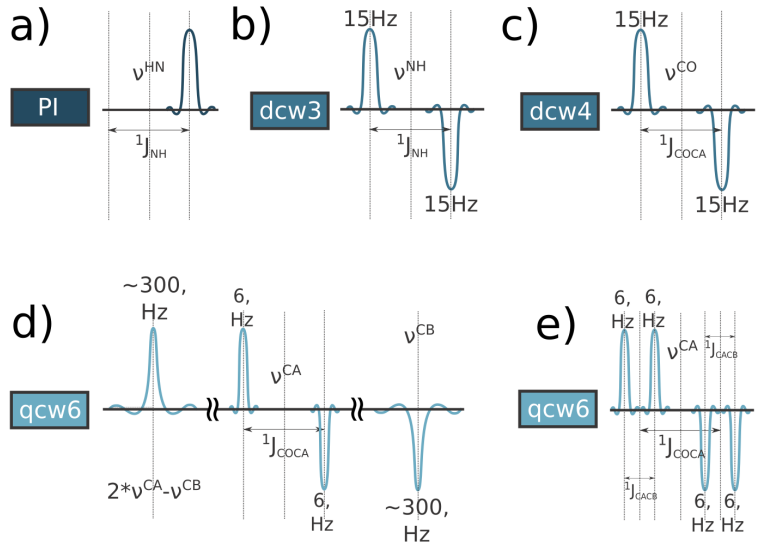

**Figure S2.** Frequency scheme of the continuous-wave sandwiches used in the different S$^4$PT modules of the FOSY experiment (Figure S1). a) *PI*: Selective single pulse for population inversion of $^1$H TROSY component for $H_z^N \rightarrow 2H_z^N N_z^H$ polarization transfer. The pulse may have a Gauss or CROP shape[8] the latter is optimized for inversion in the presence of cross-correlated relaxation.[4, 8] b) *dcw3*: Double continuous-wave sandwich includes two continuous-waves with offsets $\nu_N^i \pm \frac{J_{HN}}{2}$ starting with opposite phases x,-x (depicted as shapes pointing up/down). The B$_1$ strength for the both continuous-waves is ≈15 Hz. c) *dcw4*: Double continuous-wave sandwich includes two continuous-waves with offsets $\nu_{CO}^{i-1} \pm \frac{J_{CACO}}{2}$ ending with opposite phases ±x. The B$_1$ strength for both continuous-waves is the same ≈15 Hz. d) *qcw6*: Quadro continuous-wave sandwich qcw6 for deuterated proteins with $^{13}$C$^{B,i-1}$ decoupling includes four continuous-waves: two with $\nu_{CA}^{i-1} \pm \frac{J_{CACO}}{2}$ offsets ending with the opposite phases ±x, and two decoupling continuous-wave at $\nu_{CB}^{i-1}$ and compensation continuous-wave at $2\nu_{CA}^{i-1} - \nu_{CB}^{i-1}$ with opposite phase. The B$_1$ strength is 6 Hz for $^{13}$C$^{A,i-1}$ continuous-waves. For $^{13}$C$^{B,i-1}$ decoupling, B$_1 \gg {}^1J_{CACB}$ (e.g. B$_1$ ≈300 Hz) is selected to match/distinguish different types of amino acid. e) *qcw6*: Quadro continuous-wave sandwich for deuterated proteins without $^{13}$C$^{B,i-1}$ decoupling includes four B$_1$ = 6 Hz continuous-waves at $\nu_{CA}^{i-1} \pm \frac{J_{CACO}}{2} \pm \frac{J_{CACB}}{2}$ frequencies ending at phases x, x, -x, -x.

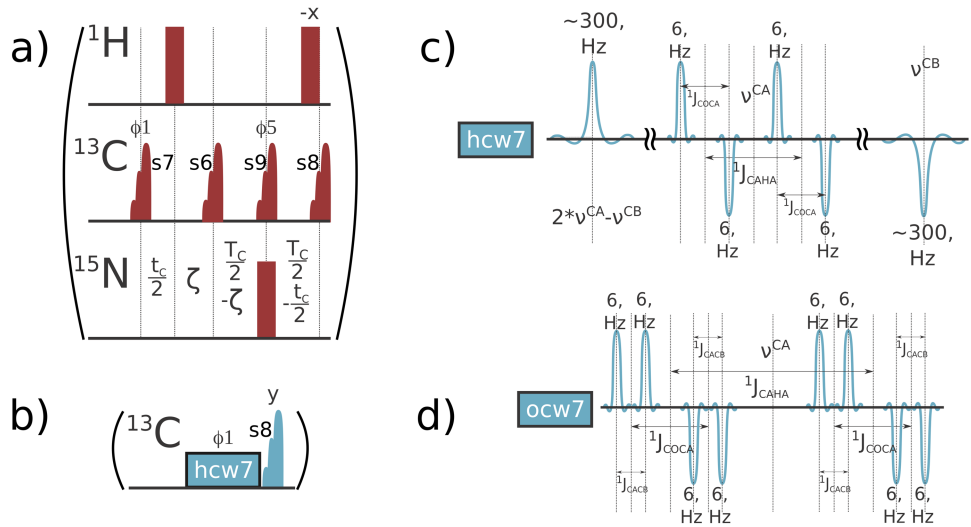

**Figure S3.** Modifications of the FOSY experiment (Figure S1) for non-deuterated proteins in H$_2$O-based buffer. The part in round brackets in Figure S1 is replaced by a block: a) For FOSY-hncoCANH (Figure S1a,b), continuous broadband $^2$H$^A$-decoupling is replaced by a pair of unselective $^1$H inversion pulses to refocus $^1J_{CAHA}$ evolution and return water and protein proton magnetization to thermal equilibrium. b) For FOSY-hncocacbNH (Figure S1c) experiment with fixed $^{13}$C$^{A,i-1}$ and selective $^{13}$C$^{B,i-1}$ decoupling, continuous $^2$H$^A$-decoupling is omitted while the quadruple $\nu_{CA}^{i-1}$ frequency selective CW irradiation module (*qcw6*) is replaced by hcw7 CW sandwich analogue. c) *hcw7*: hexo continuous-wave sandwich employing selective $^{13}$C$^{B,i-1}$ decoupling used in (b). hcw7 consists of four 6 Hz CWs at $\nu_{CA}^{i-1} \pm \frac{J_{CAHA}}{2} \pm \frac{J_{CACO}}{2}$ frequencies ending with phases x, -x, x, -x depicted as shapes pointing up (x) and down (-x), and two CWs B$_1 \gg {}^1J_{CACB}$ (e.g. B$_1$ ≈300 Hz), one for selective $^{13}$C$^{B,i-1}$ decoupling at $\nu_{CB}^{i-1}$ frequency and a compensation CW with the opposite phase at $2\nu_{CA}^{i-1} - \nu_{CB}^{i-1}$. d) *ocw7*: octo continuous-wave sandwich without $^{13}$C$^{B,i-1}$ decoupling optionally replacing *hcw7* in (b) including eight 6 Hz CWs at $\nu_{CA}^{i-1} \pm \frac{J_{CAHA}}{2} \pm \frac{J_{CACA}}{2} \pm \frac{J_{CACB}}{2}$ frequencies ending with phases x, x, -x, -x, x, x, -x, -x depicted as shapes pointing up (x) and down (-x).

## Proline – Selective Experiments

The triple resonance experiments used in this work are based on the BEST-TROSY intra-HNCA and HNcoCA experiments[9] with modifications for proline selection as described by Solyom et al.[10] Thus, a selective 8.25 ms long REBURP[11] shape $^{15}$N inversion pulse covering the distinct proline chemical shift range 138±2.77 ppm (250 Hz) is applied during the constant-time $^{13}C^A$ chemical shift evolution period ($T_C \approx 2/^1J_{CACB}$ = 56 ms). This pulse is omitted in alternating scans such that the $J_{CANpro}$ coupling either evolves or not, while the receiver phase is inverted for pairwise subtraction of scans. In the resulting spectrum, signals are only observed if the corresponding $^{13}C^A$ couples with a $^{15}$N proline spin inverted by the REBURP pulse. Thus, modified iHNCA and HNcoCA difference experiments contain signals only for residues X preceding (XP) or succeeding (PX) a proline, respectively.

## Experimental NMR technical details

All NMR experiments were recorded on an 800 MHz Bruker AVANCE IIIHD spectrometer equipped with a 3 mm TCI $^1$H/$^{13}$C/$^{15}$N cryoprobe. Spectra acquisition, processing, and analysis were performed using TopSpin 3.5 (Bruker BioSpin). 2D FOSY-hncoCA(N)H experiments were recorded with 150 × 1024 complex data points; 2D FOSY-hnco(CA)NH and 2D FOSY-hncocacbNH experiments were recorded with 65 × 1024 complex data points. The spectral widths for $^{13}$CA, $^{15}$N, and $^1$HN were 30 ppm, 26 ppm, and 14 ppm, respectively. The number of scans ranged from 2 to 32, depending on the signal intensity. The proline-selective experiments were recorded with 128 × 2048 complex data points and a $^{15}$N spectral width of 26 ppm for the PX spectrum vs. 64 × 1024 complex data points and a $^{15}$N spectral width of 41 ppm for the XP spectrum. The standard 3D BEST-TROSY HNCO experiment, obtained from the IBS library[12], was recorded with 100 × 100 × 2048 complex data points and spectral widths of 10 ppm for $^{13}$CO, 26 ppm for $^{15}$N, and 16 ppm for $^1$HN. A short recycle delay of 0.5 s was used in all experiments.

## Protein expression and purification

*E. coli* BL21(λDE3) Star™ (Novagen) cells were transformed with a modified pET28b plasmid harboring full length hTau40 protein fused to an amino-terminal His-SUMO Tag (purchased in *E. coli* codon optimized-form from GenScript). [*U*–$^{15}$N,$^{13}$C] or [*U*–$^2$H, $^{15}$N, $^{13}$C] isotope enriched protein was produced using 2xM9 minimal medium[13] supplemented with ($^{15}$N$^1$H$_4$)Cl (1 g/l) and *D*-($^1$H,$^{13}$C) or *D*-($^2$H,$^{13}$C)-glucose (2 mg/l) as the sole nitrogen and carbon sources, respectively, in either H$_2$O or D$_2$O based growth medium. All isotopically labelled material was purchased from Merck. The transformed cells were grown at 37°C in medium supplemented with 50 μg/ml kanamycin until reaching an OD$_{600}$ ≈ 0.8. Expression was induced by adding 1 mM isopropyl-thiogalactoside (IPTG) (Thermo Scientific) for 16 h at 22°C. Cells were harvested by centrifugation (5000 × g, 30min) and subsequently resuspended in lysis buffer (20 mM NaP$_i$, 500 mM NaCl, pH 7.8). Cells were lysed by four passes through an Emulsiflex C3 (Avestin) homogenizer and the cleared lysate was purified with a 5ml HisTrap HP column (GE Healthcare). hTau40 eluted within a 150 mM imidazole step. Fractions containing hTau40 were pooled and dialyzed overnight against human SenP1 cleavage buffer (20 mM TrisHCl, 150 mM NaCl, 1 mM DTT, pH 7.8). After dialysis, SenP1 protease (Addgene #16356)[14] was added and enzymatic cleavage was performed for 4 hours at room temperature. His-SUMO-Tag and hTau40 were separated by a second HisTrap HP column step and fractions containing cleaved hTau40 in the flow-through were collected, concentrated, and subsequently purified by gel filtration using a HiLoad 10/60 200pg (GE Healthcare) pre-equilibrated with NMR buffer (25 mM sodium phosphate buffer pH 6.9, 50 mM NaCl, 1 mM EDTA). The pure hTau40 fractions were concentrated to about 500 μM, flash-frozen in liquid nitrogen, and stored at -80°C till usage.

## Phosphorylation of Tau protein

GSK3β kinase was purchased from SignalChem. *In vitro* phosphorylation reactions were carried out as described before.[15] Briefly, 250 μM [*U*–$^2$H,$^{15}$N,$^{13}$C] hTau40 was mixed with 5 μl of GSK3β in phosphorylation buffer (20 mM HEPES, 2 mM ATP (Thermo Scientific), 25 mM MgSO$_4$, 2 mM EDTA, pH 7.4). The reaction was performed at 25°C for 12 hours, followed by gel filtration (Superdex® 200 Increase 10/300 GL, GE Healthcare) with NMR buffer. Under the chosen conditions, GSK3β kinase phosphorylated hTau40 to near completion (>95%), as assessed by NMR spectroscopy.

**Table S1.** Peptide sequence stretches of hTau40 conforming with the general motif (P/G)-X$^n$-p(S/T)-X$^n$-(P/G), with X ≠ (P/G), as traced out by the presented FOSY assignment approach. This list was filtered for stretches containing the phosphorylation sites reported in literature[16] and/or known kinase consensus recognition motifs. Progressing one sequence position at a time, the X amino acid can be classified as preceding a proline (-XP), following a proline (PX-), following a glycine (GX-) or none of the above, ruling out peptide sequences that cannot be mapped based on the derived sequence motif. The table shows a tick mark (√) for those residues in each stretch that conform with the amino acid type indicated by the delineated FOSY assignment protocol. Thus, starting signals 'a' and 'd' (Figure S4) could be unambiguously mapped to the phosphorylated residues pS$^{404}$ and pS$^{409}$ after only three and two FOSY steps tracing out the unique GXTp(S/T)P and PXXXp(S/T) motifs, respectively (shown bold).

| hTau40 phosphorylation sites by GSK3β[16] | Local sequence stretches | FOSY step | Start signal = 'a' | | | | Start signal = 'd' | | | |
|---|---|---|---|---|---|---|---|---|---|---|
| | | | 0 | -1 | -2 | -3 | 0 | -1 | -2 | -3 |
| | | | (S/T)P | T | X | G | (S/T) | X | X | PX |
| T175 | PAK<u>T</u>P | | √ | | √ | | | | | |
| T181 | PK<u>T</u>P | | √ | | | | | | | |
| S199 | GDRSGY<u>S</u>SP | | √ | | √ | √ | | | | |
| T205 | PG<u>T</u>P | | √ | | | | | | | |
| S212 | PGSRSR<u>T</u>P | | √ | | √ | | | | | |
| S214 | P<u>S</u>LP | | | | | | | | | |
| T217 | P<u>T</u>P | | | | | | | | | |
| T231 | PKKVAVVR<u>T</u>P | | √ | | √ | | | | | |
| S235 | PK<u>S</u>P | | √ | | | | | | | |
| S262 | G<u>S</u>TENLKHQP | | | | | | √ | | | |
| S356 | G<u>S</u>LDNITHVP | | | | | | √ | | | |
| S396 | GAEIVYK<u>S</u>P | | √ | | √ | | | | | |
| S400 | PVV<u>S</u>G | | | | | | √ | √ | | |
| **S404** | **GDT<u>S</u>P** | | √ | √ | √ | √ | | | | |
| **S409** | **PRHL<u>S</u>N** | | | | | | √ | √ | √ | √ |

**Table S2.** Chemical shift assignments obtained by the delineated FOSY approach for sequence stretch [398]VVSGDTS[404] for phosphorylated deuterated [$U$–$^2$H,$^{15}$N,$^{13}$C] p-$^2$H-hTau40, and [398]VVSGDT[403] for unphosphorylated deuterated [$U$–$^2$H,$^{15}$N,$^{13}$C] $^2$H-hTau40 and unphosphorylated protonated [$U$–$^{15}$N,$^{13}$C] $^1$H-hTau40 samples. Exact chemical shifts were obtained from the 3D BEST-TROSY HNCO[12], 2D FOSY-hnco(CA)NH, and 2D FOSY-hncoCA(N)H spectra of deuterated (Figure S1) and non-deuterated (Figure S3) hTau40. The sequential assignment walk started from the newly appearing pS[404] peak for the p-$^2$H-hTau40 sample, whereas for the unphosphorylated $^2$H-hTau40 and $^1$H-hTau40 the initial T[403] peak in HNCO spectrum was identified using the values taken from published assignment.[17]

|  |  | p-$^2$H-hTau40 | $^2$H-hTau40 | $^1$H-hTau40 |
|---|---|---|---|---|
| Ser-404 | $^1$H$^N$ | 8.43 | | |
|  | $^{15}$N$^H$ | 120.6 | | |
| Thr-403 | $^{13}$C$^O$ | 174.2 | | |
|  | $^{13}$C$^A$ | 61.1 | | |
|  | $^1$H$^N$ | 7.92 | 8.00 | 8.00 |
|  | $^{15}$N$^H$ | 114.1 | 114.7 | 114.8 |
| Asp-402 | $^{13}$C$^O$ | 176.3 | 176.6 | 176.6 |
|  | $^{13}$C$^A$ | 53.9 | 53.8 | 54.2 |
|  | $^1$H$^N$ | 8.08 | 8.06 | 8.06 |
|  | $^{15}$N$^H$ | 120.8 | 120.8 | 121.0 |
| Gly-401 | $^{13}$C$^O$ | 173.7 | 173.6 | 173.7 |
|  | $^{13}$C$^A$ | 44.7 | 44.7 | 45.1 |
|  | $^1$H$^N$ | 8.27 | 8.27 | 8.28 |
|  | $^{15}$N$^H$ | 111.6 | 111.6 | 111.8 |
| Ser-400 | $^{13}$C$^O$ | 174.8 | 174.8 | 174.8 |
|  | $^{13}$C$^A$ | 57.8 | 57.7 | 58.2 |
|  | $^1$H$^N$ | 8.27 | 8.25 | 8.26 |
|  | $^{15}$N$^H$ | 120.4 | 120.3 | 120.6 |
| Val-399 | $^{13}$C$^O$ | 176.0 | 176.0 | 176.0 |
|  | $^{13}$C$^A$ | 61.4 | 61.4 | 61.8 |
|  | $^1$H$^N$ | 8.11 | 8.12 | 8.12 |
|  | $^{15}$N$^H$ | 125.3 | 125.3 | 125.6 |
| Val-398 | $^{13}$C$^O$ | 176.2 | 176.2 | 176.2 |
|  | $^{13}$C$^A$ | 61.8 | 61.8 | 62.3 |
|  | $^1$H$^N$ | 8.00 | 8.00 | 8.01 |
|  | $^{15}$N$^H$ | 121.4 | 121.4 | 121.6 |

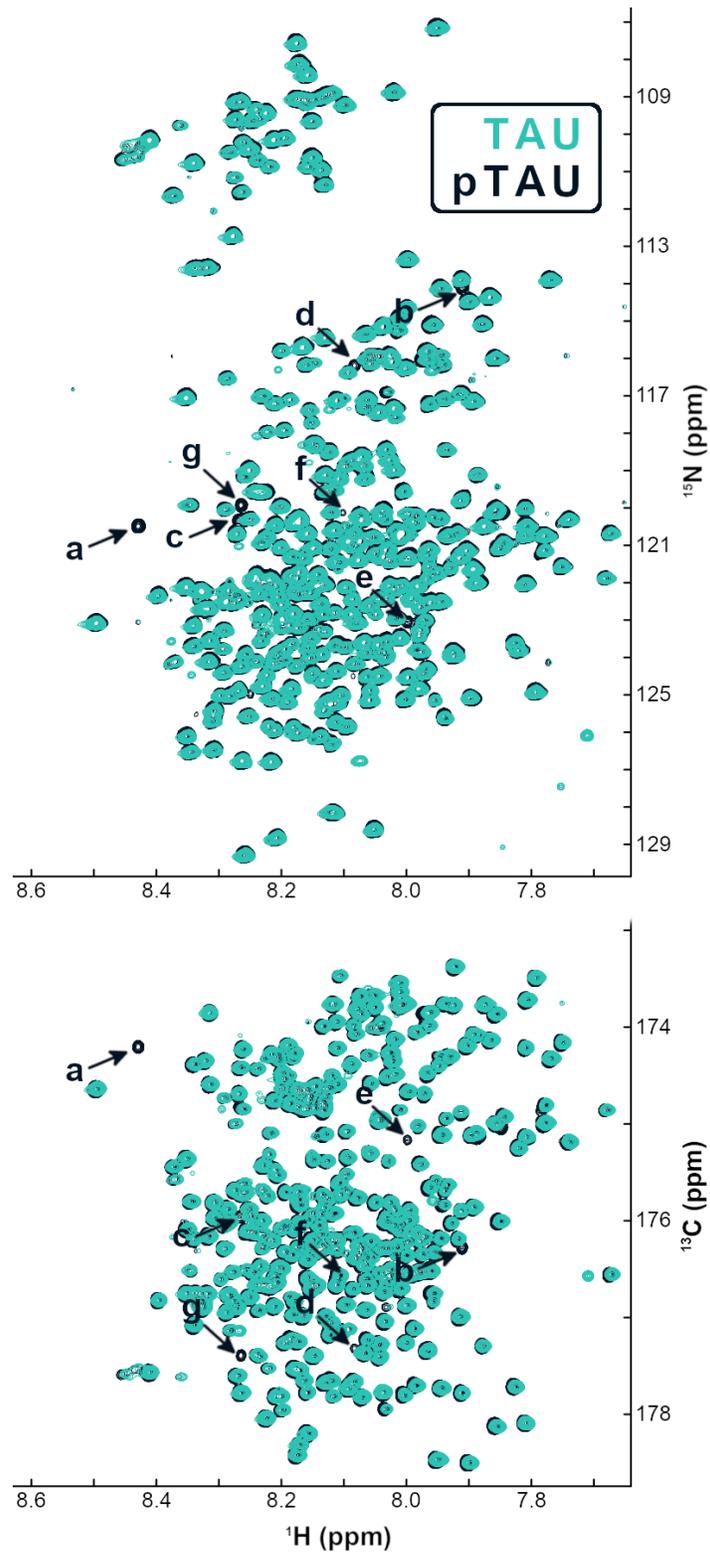

**Figure S4.** 2D $^1H,^{15}N$ (top) and $^1H,^{13}CO$ (bottom) projections of the 3D BEST-TROSY HNCO[12] spectrum of unphosphorylated hTau40 (cyan) and phosphorylated p-hTau40 (black). All identified new and shifted signals are annotated by letters 'a' to 'g'. Peaks 'a' to 'c' and 'd' to 'g' were mapped to the $pS^{404}$ and $pS^{409}$ phosphorylation sites and their preceding peptide stretches, respectively.

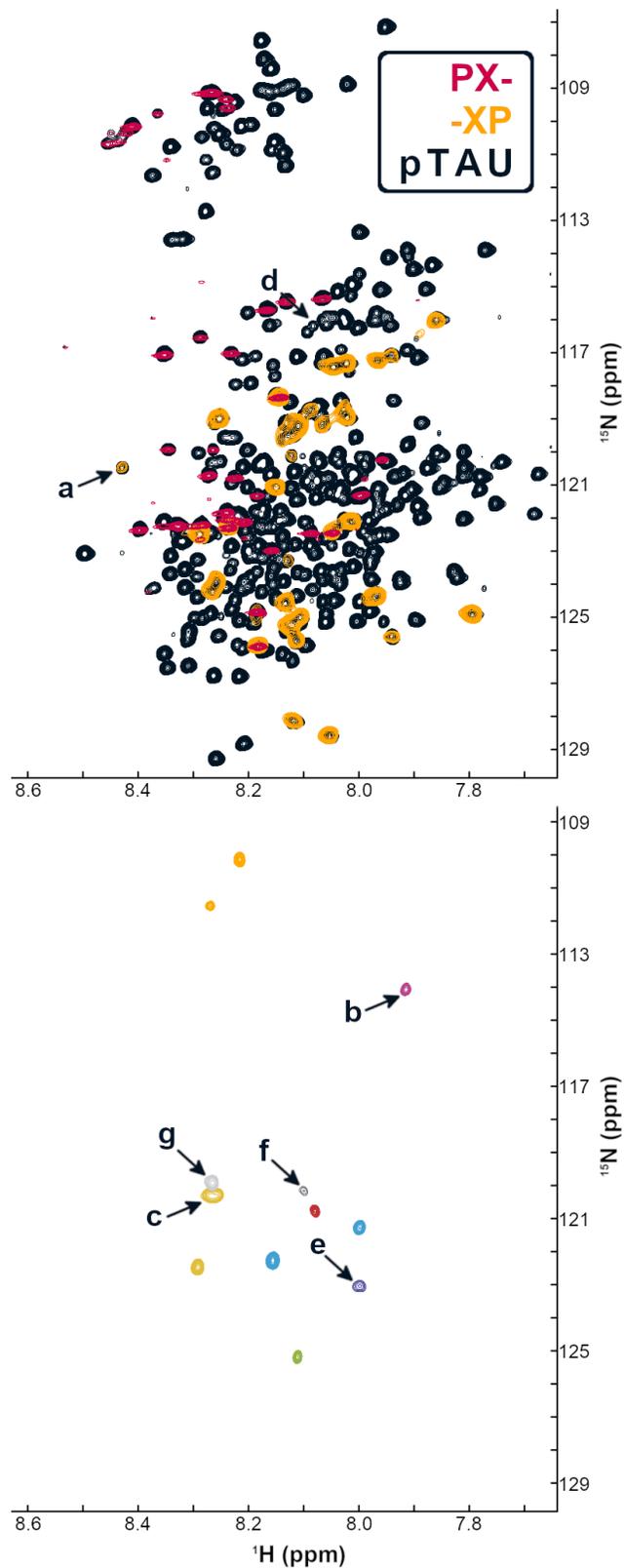

**Figure S5.** (top) Superposition of 2D $^1$H,$^{15}$N projections from the proline selective PX- (red), XP-(yellow), and regular BEST-TROSY HNCO (black) spectra of p-hTau40. The newly (after phosphorylation) appearing signals 'a' and 'd' are the starting points for local FOSY-assignment of the pertaining two phosphorylation sites. (bottom) Superposition of all nine 2D FOSY-hnco(CA)NH experiments acquired to successively correlate signals 'a' with 'b' to 'c' and 'd' with 'e' to 'g'. Each individual spectrum is coloured differently, occasionally revealing multiple peaks in a single spectrum due to overlap of the selected frequencies.

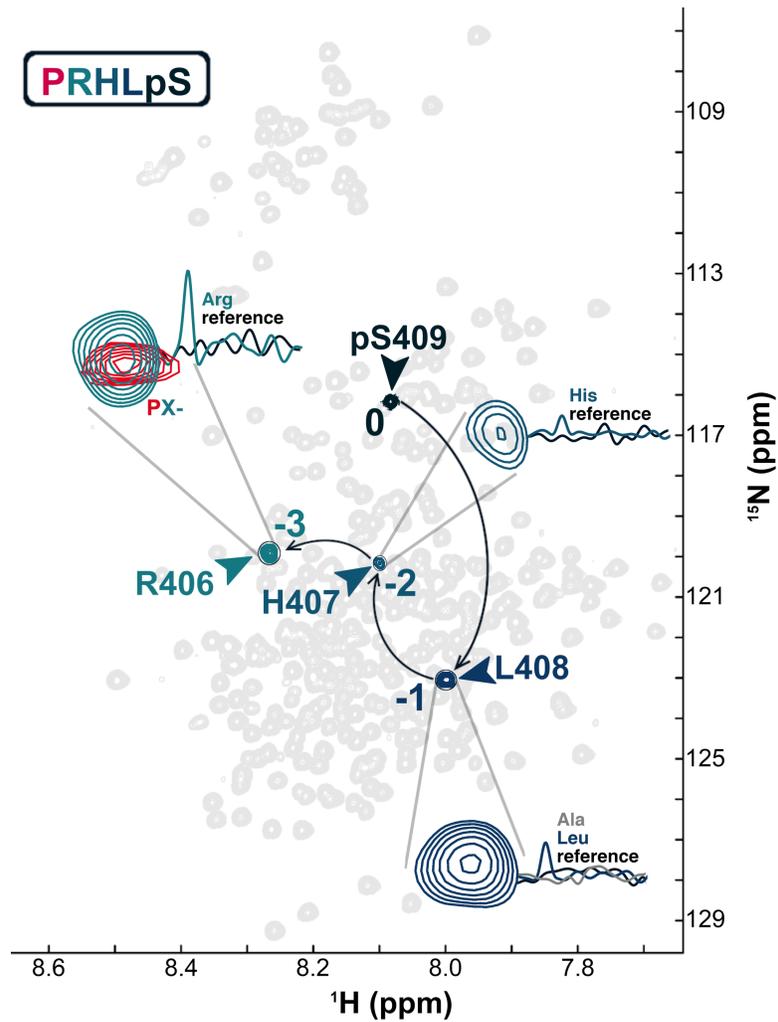

**Figure S6.** FOSY assignment walk for the second phosphorylation site in p-hTau40. The walk in the 2D $^1$H,$^{15}$N plane starts from the newly appearing signal 0 (corresponding to peak 'd' in the HNCO spectrum in Figure S4) and proceeds via the indicated three FOSY steps (-1 to -3) until reaching the signal from a PX- or GX- type residue. This traces out a PXXXp(S/T) motif that can be mapped to either PQLA$\underline{T}$[427], PVDL$\underline{S}$[316], or PRHL$\underline{S}$[409] stretches in the hTau40 primary sequence. FOSY-hncocacbNH is then used to narrow down on the pertaining amino acid type, indicating a Leu (not Ala) for signal -1, His for signal -2, and Arg for signal -3. Thus, PRHL$\underline{S}$[409] is unambiguously confirmed as correct assignment. This stretch is also the only one of the three alternatives that was shortlisted as a known GSK3β-mediated phosphorylation site (Table S1).